\pdfoutput=1

\documentclass[11pt]{article} 

\usepackage[utf8]{inputenc} 
\usepackage{geometry} 
\usepackage{graphicx} 
\usepackage{booktabs} 
\usepackage{array} 
\usepackage{paralist} 
\usepackage{verbatim} 
\usepackage{subfig} 
\usepackage{fancyhdr} 
\usepackage{sectsty}
\usepackage[nottoc,notlof,notlot]{tocbibind} 
\usepackage[titles,subfigure]{tocloft} 

\usepackage{amsmath}
\usepackage{amssymb}
\usepackage{amsthm}
\usepackage[usenames, dvipsnames]{color}

\allsectionsfont{\sffamily\mdseries\upshape} 

\newtheorem{theorem}{Theorem}

\title{A fresh look at igorability for likelihood inference}
\author{JC Galati$^1$}
\date{%
    \today \\
    \mbox{}\\%
    $^1$\small{Department of Mathematics and Statistics, La Trobe University, Melbourne, VIC 3083}
}

\begin{document}
\maketitle

\begin{abstract}
   When data are incomplete, a random vector $Y$ for the data process together with a binary random vector $R$
   for the process that causes missing data, are modelled jointly. We review conditions under which $R$ can be
   ignored for drawing likelihood inferences about the distribution for~$Y$. The standard approach of Rubin\;(1976)
   and Seaman\;et.\,al.\,(2013),\,\textit{Statist.\,Sci.},\,\textbf{28}:2\;pp.\,257--268 emulates complete-data methods
   exactly, and directs an investigator to choose a full model in which missing at random (MAR) and distinct of
   parameters holds if the goal is not to use a full model. Another interpretation of ignorability lurking in the literature
   considers ignorable likelihood estimation independently of any model for the conditional distribution
   $R$ given~$Y$. We discuss shortcomings of the standard approach, and argue that the  alternative gives the
   `right' conditions for ignorability because it treats the problem on its merits, rather than emulating methodology
   developed for when the investigator is in possession of all of the data.
  
   \vspace*{2mm}
\end{abstract}

\section{Introduction} \label{Sect:Intro}
Missing data are a common problem in empirical research, and particularly so in medical and epidemiological studies.
A central feature of the statistical methods for dealing with incomplete data pertains to conditions under which the
random vector for the process causing the missing data need not be modelled. The modern framework was
introduced by Rubin\;(1976). If $Y$ is a random vector representing the data generation process, Rubin\;(1976)
introduced the concept of modelling missingness through a corresponding binary random vector $R$ of the same
dimension as~$Y$, together with a joint probability distribution for $(Y, R)$. The realisations of $Y$ comprise
complete data, both observed and unobserved, and realisations of $R$ determine which values of $Y$ are observed
and which are missing.  The conditional distribution for $R$ given~$Y$ represents the process that causes missing
data, herafter called the \textbf{missingness process}.

Given a model of joint densities for $(Y, R)$, Rubin (1976) identified conditions on the model for which the same
inferences result whether the full model is used or the model for the missingness proces is discarded. Rubin considered
direct likelihood, Bayesian and sampling distribution paradigms, but not frequentist likelihood inference specifically.
Seaman\;et.\,al.\,(2013) reviewed use of the ignorability conditions in the literature to promote unity amongst writers,
and adapted Rubin's conditions specifically for frequentist likelihood inference. We refer to the conditions derived in
these works as the \textbf{standard conditions} for ignorabiliy.

Seaman\;et.\,al.\,(2013,\,p.\,266) identified an alternative interpretation of ignorabilty lurking in the literature. This
approach treats ignorable likelihood as an estimation process in its own right, independently of any model for the
missingness process. Our aim is to review the two approaches, to explain some shortcomings of the standard
approach, and to argue that the alternative interpretation gives the `right' conditions for ignorability because it treats
the problem on its merits, rather than adopting methodology developed for when the  investigator is in possession of
all of the data.

\section{Ignorability for direct likelihood inference (Rubin,\,1976)} \label{Sect:Rubin}
We retain the notation $Y$\;and\;$R$ from Section~\ref{Sect:Intro}. The starting point for ignorability in the sense
of Rubin (1976) is a (full) model of joint densities for $(Y, R)$:
\begin{equation}
   \mathcal{M}_g \,=\, \{\, f_\theta(\mathbf{y})\,g_\psi(\mathbf{r} |\, \mathbf{y}) : (\theta, \psi)\in\Delta \,\}.
   \label{Eq:MgModel}
\end{equation}
If $\Theta$ and $\Psi$ are the images of the projections $(\theta,\psi)\mapsto\theta$ and $(\theta,\psi)\mapsto\psi$,
respectively, then the \textbf{data model} of $\mathcal{M}_g$ is
\begin{equation}
   \mathcal{M}_s \,=\, \{\, f_\theta(\mathbf{y}) : \theta\in\Theta \,\}
   \label{Eq:MsModel}
\end{equation}
and the \textbf{missingness model} is $\{\,g_\psi(\mathbf{r}|\,\mathbf{y})\,:\,\psi\in\Psi\,\}$. We call each density
in the missingness model a \textbf{missingness mechanism}. Recall from Section~\ref{Sect:Intro} that the two
conditions under which the missingness model could be discarded and identical direct likelihood inferences drawn
from a model for the observed data derived from\;(\ref{Eq:MsModel}) were called MAR and distinctness of
parameters. We consider these in turn.

Given a realisation $(\mathbf{y}, \mathbf{r})$ of the random vector~$(Y, R)$, let~$\mathbf{y}^{ob(\mathbf{r})}$
and~$\mathbf{y}^{mi(\mathbf{r})}$ denote the vectors of observed and missing components of~$\mathbf{y}$,
respectively. Analysis of the observed data $(\mathbf{y}^{ob(\mathbf{r})}, \mathbf{r})$ then proceeds by
restriction of the random vector $(Y, R)$ to the event
\begin{equation} 
   \{\, (\mathbf{y}_*, \mathbf{r}_*) \,:\,
      \mathbf{r}_* = \mathbf{r} \text{ and }
      \mathbf{y}_*^{ob(\mathbf{r})} = \mathbf{y}^{ob(\mathbf{r})} \,\}
   \label{Eq:ODE}
\end{equation}
comprising all datasets $\mathbf{y}_*$ (together with~$\mathbf{r}$) which correspond
to~$(\mathbf{y}, \mathbf{r})$ on the observed data values~$\mathbf{y}^{ob(\mathbf{r})}$ but may differ on
the unobserved values~$\mathbf{y}^{mi(\mathbf{r})}$. A missingness mechanism
$g_\psi(\mathbf{r} |\, \mathbf{y})$ is called \textbf{missing at random} (MAR) with respect
to~$(\mathbf{y}, \mathbf{r})$ if $g_\psi(\mathbf{r} |\, \mathbf{y})$ is a constant function on the
set~(\ref{Eq:ODE}), where $g$ is considered to be a function of $\mathbf{y}$ with $\mathbf{r}$ held fixed.
Rubin (1976) defined missing at random to be a property of the full model (\ref{Eq:MgModel}) by requiring each
density in the model to be MAR with respect to~$(\mathbf{y}, \mathbf{r})$. The terminology \textbf{realised MAR}
was introduced in Seaman\;et.\,al.\,(2013) to distinguish this weaker form of MAR from a stronger form more suited
to frequentist likelihood inference: a missingness mechanism is called \textbf{everywhere MAR} if it is realised MAR
with respect to all possible data vectors and response pattens, $(\mathbf{y}, \mathbf{r})$, not just the realised pair
representing the observed and missing data.

The second of Rubin's conditions, \textbf{distinctness of parameters}, requires the parameter space $\Delta$
of~(\ref{Eq:MgModel}) to be a direct product $\Delta = \Theta\times\Psi$ of parameter spaces $\Theta$ of the
data model and the missingness model.

If every missingness mechanism in (\ref{Eq:MgModel}) is realised MAR with respect to the realised data
vector~$\mathbf{y}$ and response pattern~$\mathbf{r}$, then the likelihood function for that part of
(\ref{Eq:MgModel}) pertaining to the observable data factorizes as
\begin{equation}
   L_g(\theta, \psi) \,=\,
      \int f_\theta(\mathbf{y}) \, g_\psi(\mathbf{r} |\, \mathbf{y}) \,
      \text{d}\mathbf{y}^{mi(\mathbf{r})} \,=\,
      g_\psi(\mathbf{r} |\, \mathbf{y}) \, \int f_\theta(\mathbf{y}) \, \text{d}\mathbf{y}^{mi(\mathbf{r})}\,.
   \label{Eq:MgLikelihood}
\end{equation}
If, in addition, distinctness of parameters holds for (\ref{Eq:MgModel}), then the (maximal) domain of each mapping
$\theta\mapsto L_g(\theta,\psi)$ is the same for every value of~$\psi$, and likelihood estimates for $\theta$ can be
obtained by maximising the simpler function
\begin{equation}
   L_s(\theta) \,=\,
      \int f_\theta(\mathbf{y})\,
      \text{d}\mathbf{y}^{mt(\mathbf{r})}\,.
   \label{Eq:MsLikelihood}
\end{equation}
over its full domain. We refer the reader to Rubin (1976), Seaman\;et.\,al.\,(2013) and Mealli and Rubin (2015) for
additional details.

\vspace*{3mm}
\textbf{An aside.} In (\ref{Eq:MsLikelihood}) we have used $\mathbf{y}^{mt(\mathbf{r})}$ instead of
$\mathbf{y}^{mi(\mathbf{r})}$ to denote the unobserved variables because overlaying the response pattern
$\mathbf{r}$ onto the marginal distribution for $Y$ involves a different relationship between $R$ and $Y$ compared
to $(Y, R)$. The former does not respect the stochastic relationship encoded in the random vector~$(Y, R)$ because
it involves holding $R$ fixed and allowing the marginal distribution for~$Y$ to vary. The `t' in
$\mathbf{y}^{mt(\mathbf{r})}$ can be interpreted to mean `these are the variables of the marginal distribution for
$Y$ that were missing this time.' Note also that we do \textbf{not} do the same on the right hand side of
(\ref{Eq:MgLikelihood}) because the set being integrated over is all of $R\times Y,$ whereas in
(\ref{Eq:MsLikelihood}) it is only~$Y$. See Galati (2019) for further details.
\hfill $\qedsymbol$

\vspace*{3mm}
It is helpful to view the two ignorability conditions, MAR and distinctness of parameters, in a hierarchy as follows:
\begin{itemize}
   \item[(\textbf{a})]
      Does the investigator wish to enforce a relationship $\Delta\subsetneq\Theta\times\Psi$ between $\theta$
      and $\psi$ for models (\ref{Eq:MgModel}) when estimating~$\theta$?  If so, the analyst has no option but to
      consider only full models  for which distinctness of parameters \textbf{does not} hold, irrespective of whether
      of not densities in the model are realised MAR.

   \item[(\textbf{b})]
      If the answer to (a) is no, then is every missingness mechanism in the model (\ref{Eq:MgModel}) realised MAR? If
      so, the analyst can discard the missingness mechanisms from the model. 
\end{itemize}
Viewed in this way, ignorability for direct likelihood inferences is seen to be comprised of two components, the
distinctness of parameters criterion, which really is just a statement that the investigator has no relationship of the form
$\Delta\subsetneq\Theta\times\Psi$ to enforce when estimating~$\theta$, and the MAR component, which establishes
the relationship between (\ref{Eq:MgLikelihood}) and (\ref{Eq:MsLikelihood}) for fixed~$\psi$.

Note also this does \textbf{not} mean that the MAR condition is of no use when a relationship of the form
$\Delta\subsetneq\Theta\times\Psi$ \textbf{is} enforced (that is, when distinctness of parameters does \textbf{not}
hold). In this case, the MAR condition allows a likelihood $L(\psi)=g_\psi(\mathbf{r}|\,\mathbf{y})$ for the
missingness model to be maximised independently of~$\theta$, and then $\Delta$ can be used to determine an
appropriate restriction on the domain of (\ref{Eq:MsLikelihood}) for estimating~$\theta$.

We will call full models (\ref{Eq:MgModel}) satifying the distinctness of parameters and MAR criteria 
\textbf{ignorable models}, and we emphasise that Rubin's (1976) ignorability theory for direct likelihood inference
identifies a subset of models of the form (\ref{Eq:MgModel}), the ignorable models, from which an investigator can
choose their model if they wish to draw inferences for $\theta$ free from the inconvenience of needing to model the
missingness process explicitly.

\section{Limitations caused by missing data} \label{Sect:Limitations}

Ignorability is often presented as having something to do with drawing valid inferences. For example,
Rubin\;(1976,\;Summary) states that the ignorability conditions are ``\textit{the weakest general conditions under
which ignoring the process that causes missing data always leads to correct inferences.''} `Correct inferences' in this
instance seems to mean that inferences will be drawn from the correct likelihood given the chosen model. It has
nothing to do with whether or not the choice of model is valid for the given data, or whether or not valid conclusions
will be drawn from the data.

In the model-based paradigms, validity in the latter sense mentioned above is a subjective assessment of the goodness
of fit of the model to the data. If the model fits poorly, then in some sense the inferences are not justifiable, and if the
model fits too-well, then the model becomes more a description of the specific realised dataset rather than a description
of the process which generated the data.

When data are incomplete, the philosophy of the model-based likelihood paradigm breaks down in two essential
ways. Firstly, it is impossible to validate the investigator's choice of missingness model against the data because the
data required for this are missing (Molenberghs\;et.\,al.\,(2008)). So consideration of a missingness model becomes
hypothetical in a manner analogous to the frequentist paradigm's hypothetical assumptions about $(Y, R)$. In the
literature, this feature of incomplete data methods typically is referred to as `untestable assumptions.'

The second way in which the paradigm breaks down is that it becomes impossible to validate even a model for the
observed data against the observed data. The reason for this is a little more subtle.  If the possible missingness
patterns realisable from $R$ are $\mathbf{r}_1, \mathbf{r}_2, \ldots, \mathbf{r}_k$, and these occur with
marginal probabilities $p_1, p_2, \ldots, p_k$, then a density $f(\mathbf{y})$ for the marginal distribution for $Y$
can be written as the mixture
\begin{equation} 
   f(\mathbf{y}) \;=\; \sum_{i=1}^k\;p_i\,p(\mathbf{y}|\,\mathbf{r}_i)
   \label{Eq:PatternMixture}
\end{equation}
where $p(\mathbf{y}|\,\mathbf{r}_i)$ is the conditional density for $Y$ given $R=\mathbf{r}_i$. If
$\mathbf{r}_1=(1,1,\ldots,1)$ is the pattern for a complete case, then $p(\mathbf{y}|\,\mathbf{r}_1)$ gives
the distribution for the complete cases, which may differ from the marginal distribution for $Y$ given
by~$f(\mathbf{y})$. And in general, for the $i^{th}$ missingness pattern~$\mathbf{r}_i$, the distribution of
the $\mathbf{y}$ values realised with $\mathbf{r}_i$ is $p(\mathbf{y}|\,\mathbf{r}_i)$ and \textbf{not}
$f(\mathbf{y})$. Additionally, the distribution for the \textit{observed} values realised with $\mathbf{r}_i$ is
given by the marginal density $\int\,p(\mathbf{y}|\,\mathbf{r}_i)\,d\mathbf{y}^{mi(\mathbf{r}_i)}$. This
marginalisation stratifies the distribution for $Y$ into pieces of different shapes such that the complete data
underlying each shape typically is not distributed according to~$f(\mathbf{y})$, and the differing shapes makes it
impossible to mix them back together to recover $f(\mathbf{y})$ via~(\ref{Eq:PatternMixture}). The result is that
the observed data comprise a collection of subsamples from different distributions, no one of which can be used to
assess the fit of the data model, and the irregular shapes prevent the subsamples from being pooled together.

To overcome the difficulties associated with checking the fit of the data model to the observed data, we note that
imputation-based methods combined with posterier predictive checks have been considered, but we do not elaborate
on these techniques. The points we wished to make are summarised below:
\begin{itemize}
   \item[(\textbf{c})]
      When carrying over a model-based philosophy to the case of incomplete data, the types of hypothetical
      considerations typically rejected by these philosophies become inescapable due to the impossibility of validating
      the fit of the missingness model to the observed data.

   \item[(\textbf{d})]
      Validating the fit of the data model to the observed data becomes substantially more complicated when the data
      are incomplete.
\end{itemize}

\section{The middle road: When $R$ is MAR} \label{Sect:RisMAR}
With any data analysis, the intention typically is to model the data to answer some substantive question under
investigation. But incompleteness of the data impedes analysis in two ways, the dataset has an irregular shape, and the
underlying missingness process can distort the distribution of the data that the investigator can observe. Methods for
taking these factors into account differ in their difficulty and inconvenience, and as a matter of practicality, often
methods that are less inconvenient are accorded priority ahead of more difficult and inconvienient methods. 

Methods like multiple imputation overcome the irregular shape of the data by filling in missing values with plausible
values, and adjusting the precision of estimates accordingly (Molenberghs\;et.\,al.\,2015). A first step in creating
imputations is often to consider a model of the form (\ref{Eq:MsModel}) to model jointly the variables in the dataset.
To estimate~$\theta$, likelihood estimation might be employed. However, this estimation is further impeded by the
potential distorting effect of the missingness process on the distribution of the observable data. The choice the analyst
has at hand is to use ignorable likelihood estimation based on (\ref{Eq:MsLikelihood}) anyway, or to model the
missingness process and base estimation of $\theta$ on the left hand side of~(\ref{Eq:MgLikelihood}). Apart from
the class of ignorable models, the latter adds a substantial layer of complexity and inconvenience, and understanding
conditions under which this can be avoided is important.

In the situation just described, the primary concern is not in understanding the conditions under which modelling the
missingness process would be unnecessary, as Rubin (1976) considered. Rather, the primary concern is simply to
obtain an estimate for~$\theta$, and to understand the conditions under which this can be done without the need to
consider models for the missingness process at all. This question cannot be answered using the approach in
Rubin\;(1976), reviewed in Section~\ref{Sect:Rubin}, because no model for the missingness process is posited against
which to compare the estimates from the ignorable likelihood estimation.

Seaman\;et.\,al.\,(2013, p\,266) note that it is this question that some writers seem to have taken as their interpretation
of ignorability. While these writers were considering frequentist properties of estimation, the same ideas apply to direct
likelihood inferences. We elaborate on the details. Suppose that
\begin{equation}
   h(\mathbf{y}, \mathbf{r}) = f(\mathbf{y})g(\mathbf{r}|\,\mathbf{y})
   \label{Eq:FullDensity}
\end{equation}
is a joint density for the random vector~$(Y, R)$, and consider the model 
\begin{equation}
   \mathcal{M}_t \,=\, \{\, f_\theta(\mathbf{y})\,g(\mathbf{r} |\, \mathbf{y}) : \theta\in\Theta \,\}.
   \label{Eq:MtModel}
\end{equation}
By definition (\ref{Eq:MtModel}) is correctly specified for the missingness process. We can ask under what conditions
can the likelihood for the observable data for this model,
\begin{equation}
   L_t(\theta) \,=\,
      \int f_\theta(\mathbf{y}) \, g(\mathbf{r} |\, \mathbf{y}) \,
      \text{d}\mathbf{y}^{mi(\mathbf{r})}
   \label{Eq:MtLikelihoodNotMAR}
\end{equation}
be maximised without needing to evaluate the unknown density function $g(\mathbf{r}|\,\mathbf{y})$? If
$g(\mathbf{r} |\, \mathbf{y})$ is MAR with respect to the realised values~$(\mathbf{y}, \mathbf{r})$,
then~(\ref{Eq:MtLikelihoodNotMAR}) factorises in the usual way
\begin{equation}
   L_t(\theta) \,=\,
       g(\mathbf{r} |\, \mathbf{y}) \, \int f_\theta(\mathbf{y})\,
      \text{d}\mathbf{y}^{mi(\mathbf{r})}
   \label{Eq:MtLikelihoodMAR}
\end{equation}
and~(\ref{Eq:MtLikelihoodMAR}) can be maximised without needing to evaluate~$g(\mathbf{r} |\, \mathbf{y})$.
Therefore, under this MAR assumption about~$g(\mathbf{r} |\, \mathbf{y})$, maximising
(\ref{Eq:MtLikelihoodMAR}) is equivalent to maximising~(\ref{Eq:MsLikelihood}). Hence, if the investigator would
have no reason to impose a relationship $\Delta\subsetneq\Theta\times\Psi$ on the parameters of a
model~(\ref{Eq:MgModel}), if such a model were to be considered, and if the investigator is happy to assert that the
missingness process itself is realised MAR, then direct likelihood inferences for $\theta$ can be obtained by ignorable
likelihood estimation without the need to consider a model for the missingness process at all. In particular, considering
some hypothetical model (\ref{Eq:MgModel}) and asserting distinctness of parameters (that is, choosing an ignorable
model as the starting point) is simply unnecessary. We record this formally.

\begin{theorem}[Missingness-model-free Ignorability] \label{Thm:MMFIgnorabilty}
   If the investigator would have no reason to impose a relationship $\Delta\subsetneq\Theta\times\Psi$ on the
   parameters of a model~(\ref{Eq:MgModel}), and if the distribution for the random vector $R$ conditional on~$Y$
   is realised MAR, then there is no need to consider models of the form (1) at all. In this case, direct likelihood inferences
   for $\theta$ can be obtained by ignorable likelihood estimation, and this equates to the investigator using the
   (unknown) conditional distribution for $R$ given $Y$ directly in the analysis.
   \hfill $\qedsymbol$
\end{theorem}

When interpreting ignorability in this `model-free' sense, writers typically have gone further and adopted a frequentist
view in which ignorable likelihood estimation retains the asymptotic properties of likelihood theory
(Seaman\;et.\,al.\,2013,\;p.\,266). For completeness, we review the conditions that would be needed for ignorability in
this sense.

Recall from Section~\ref{Sect:Rubin} that ignorability for likelihood inferences in the sense of Rubin (1976) has two
facets, with one being whether or not the investigator wishes to impose e a relationship
$\Delta\subsetneq\Theta\times\Psi$ onto the estimation of~$\theta$. This consideration applies irrespective of the mode
of likelihood inference, and we retain consideration of non-distinctness of parameters as the first step in the decision
making process. When this would be of no interest to the investigator, a correctly specified ignorable model together
with Theorem~\ref{Thm:MMFIgnorabilty} implies the investigator can dispense with consideration of the ignorable
model, and instead assert directly that $R$ given $Y$ is MAR.

For reasons explained in Seaman\;et.\,al.\,(2013), to consider frequentist properties of likelihood theory, we
strengthen our assumption to $R$ given $Y$ being everywhere MAR. This is then sufficient for ignorable likelihood
estimation to be valid in the frequentisti sense of likelihood theory provided the additional hypotheses of likelihood
theory are satisfied. These requirements are summarised below.

The model for the observable data is obtained by removing from each vector in $Y\times R$
the coordinates pertaining to the missing values. This creates an irregularly-shaped set. The probability measure on
this irregularly-shaped set is obtained by pulling back events in this set to unions of events of the form
(\ref{Eq:ODE}) in~$Y\times R$, and integrating the densities in (\ref{Eq:MtModel}) over these corresponding
events for~$(Y, R)$ (by applying iterated integrals as per Fubini's Theorem {(Ash and Dol\'{e}ans-Dade 2000,\,p.\,101))}.
In this way, the functions on the right had side of~(\ref{Eq:MtLikelihoodNotMAR}) are seen to give a model of densities
for the observable data.

By constuction, $\mathcal{M}_t$ is correctly specified if, and only~if, $f(\mathbf{y})\in\mathcal{M}_s$.
The integration in~(\ref{Eq:MtLikelihoodNotMAR}) sets up a mapping from $\mathcal{M}_t$ to the observable data
model. The observable data model therefore will be correctly specified if, and only if, $\mathcal{M}_t$ is, and
identifiable provided $\mathcal{M}_s$ is identifiable (different values of $\theta$ correspond to different density
functions $f_\theta$) and the mapping to it from $\mathcal{M}_t$ is one-to-one. A sufficient condition for the latter to
be true is that the missingness process assigns non-zero probability to complete cases for all values of~$\mathbf{y}$,
since then densities~(\ref{Eq:MtLikelihoodMAR}) corresponding to different values of $\theta$ will disagree for at
least one $\mathbf{y}$ value on that part of the model pertaining to the complete cases. Finally, the appropriate
regularity conditions must be satisfied by $L_t(\theta)$, $\Theta$ and the value $\theta_0\in\Theta$ for which
$f_{\theta_0}=f$.

We summarize this formally as follows.
\begin{theorem}[Ignorability for frequentist likelihood inference]
   Sufficient conditions for ignoring the missingness process when drawing frequentist likelihood inferences are:
   \begin{enumerate}
      \item there is no relationship $\Delta\subsetneq\Theta\times\Psi$ (see (\ref{Eq:MgModel})) to be imposed on
         the analysis,

      \item \label{Item:MAR} the distribution of $R$ given $Y$ is everywhere MAR,

      \item the additional requirements of likelihood theory (summarised above) are satisfied.

   \end{enumerate}
   Moreover, when condition\;\ref{Item:MAR} holds, ignorable likelihood estimation equates to using the (unknown)
   distribution for $R$ given $Y$ directly in the analysis. In these circumstances, ignoring the missingness process is
   preferable to modelling it explicitly.
   \hfill $\qedsymbol$
\end{theorem}

\section{Discussion} \label{Sect:Discussion}
The foundations for ignorability of the process that causes missing data were put in place by Rubin (1976). With the
exception of a stronger form of MAR framed in Seaman\;et.\,al.\,(2013) for frequentist likelihood theory, these
 foundations have been accepted essentially unaltered for more than four decades. Despite this, the conditions for
ignorability seem to be more confusing than they should be.

One reason for this might be that Rubin (1976) presented distinctness of parameters as a mathematical requirement of
ignorable likelihood estimation. We suggest that this criterian is better understood from a statitistical perspective,
namely, whether or not the investigators wish to impose on the analysis a relationship
$\Delta\subsetneq\Omega\times\Psi$ between the parameters for the data densities and the missingness mechanisms.
From this perspective, non-distinctness of parameters is a choice of restriction incorporated in the analysis by the
investigator, not a mathematical requirement that makes ignorable estimation `work.'  Moreover, the situations in
which imposing such a restriction could be considered reasonable would seem to be rare. In many cases, problems
similar to defining statistical significance to be~`$p<0.05$' would arise. We suggest that the condition should be
expressed as non-distinctness of parameters, and that it should serve more as a footnote to the theory, rather than
being given such a prominant place.

Another factor which may be contributing to confusion about the concept is the linking of ignorability with notions
of `valid' inferences. For example, Rubin (1976, Summary) communicates the implications of these conditions as
``\textit{always leads to correct inferences},'' while Little and Rubin (2002,\;p\,120) refer to inferences being
``\textit{valid from the frequency perspective},'' and Seaman\;et.\,al.\,(2013, Abstract) simply refer to
``\textit{valid inference}.'' In these cases, what the terms mean is left undefined. However, linking ignorability
with validity of inferences at all would seem to be highly misleading because ignorability is silent on whether or not a
particular choice of missingness model is a `valid' choice for the data at hand, and it is equally silent on whether the
chosen data model is `valid' for the data at hand. So it is difficult to see that there can be any meaningful sense in
which satisfaction of the ignorability conditions implies that inferences drawn from the model will be valid.

We suggest, however, that a primary source of confusion surrounding ignorability is likely to be because, as
framed, it is derived by emulating complete-data methods without modification. Specifically, a full model for $(Y, R)$
is taken as the starting point for the model-based paradigms, and correct specification of the full model is added for
frequentist likelihood inference. These paradigms are predicated on the investigator being in possession of the data
to enable validation of the model against the data, and this is not the case when data are incomplete. As discussed
in Section~\ref{Sect:Limitations}, these complete-data paradigms do not carry over completely to incomplete data
because the model for the missingness process cannot be validated against the data. This feature of the incomplete
data setting undermines the rationale for considering a model for the missingness process as the starting point for
an analysis.

Additionally, by framing ignorability in terms of properties of the model for the missingness process, instead of in
terms of the missingness process itself, the usual causal link between choice of model and properties of estimation is
partially severed. Changes to the missingness model can be made without altering the ignorable likelihood estimator
in any way at all. While it is true that swapping one ignorable missingness model for another merely results in a
proportional change in the likelihood, this forces a user of the tools to be in possession of unnecessary detail about
the relationship between the estimation process and some `hypothetical'  set of models for the missingness process.

Possibly the strongest argument against the standard conditions is the convoluted nature of the way the question
posed is answered. Specifically, for an investigator to choose \textbf{not to use} a full model, the investigator is 
directed \textbf{to use} a full model with specific properties, when making any choice of full model at all is
unnecessary.
  
The alternative interpretation reviewed and fleshed out in Section~\ref{Sect:RisMAR} avoids these issues by making
an assumption directly about the conditional distribution $R$ given~$Y$. This has no analogue in the correponding
complete data methods (which do not make direct assumptions about the data random vector~$Y$). However,
doing so is the most direct and natural way to answer the ignorability question: the two scenarios under which a full
model is required are (i) $R$ given $Y$ is not MAR, or (ii) the investigator has prior information about a relationship
between the data distribution and missingness process to incorporate into the estimation of~$\theta$. Otherwise,
ignorable estimation is appropriate because it equates to (the unknown) $R$ given $Y$ being used directly in the
analysis.

Proponents of model-based paradigms might argue that the properties of $(Y, R)$ can never be known in reality.
While it is true that the MAR assumption is `hypothetical' (untestable), ascribing it to a model for $R$ given~$Y$
has no advantages over ascribing it directly to $R$ given~$Y$ because it is impossible to validate this property for
the missingness model against the data. In short, choosing an untestable model is not an improvement on making an
untestable assumption, and comes with the disadvantages discussed above.

In relation to the ignorable models identified by Rubin (1976), the primary difference between the standard conditions
and the alternative interpretation of ignorability reviewed in Section~\ref{Sect:RisMAR} can be summed up as follows:
in the former case, the ignorable models are the full models that an investigator would choose from in order to not use
a full model; in the latter case, the ignorable models are the full models that can be ignored because the need for an
investigator to contemplate ever choosing one never arises.


\end{document}